\documentclass[conference]{IEEEtran}
\IEEEoverridecommandlockouts

\usepackage{graphicx}
\usepackage{multirow}
\usepackage{subfigure}
\usepackage[noadjust]{cite}
\newcommand{\RNum}[1]{\uppercase\expandafter{\romannumeral #1\relax}}
\usepackage{physics}
\usepackage{placeins}  % add to preamble
\usepackage{array}
\usepackage{booktabs} 
\newcommand{\ab}{\alpha\beta}
\newcommand{\mbf}[1]{\mathbf{#1}}

\ifCLASSINFOpdf
\else
\fi
\begin{document}
%
% paper title
% Titles are generally capitalized except for words such as a, an, and, as,
% at, but, by, for, in, nor, of, on, or, the, to and up, which are usually
% not capitalized unless they are the first or last word of the title.
% Linebreaks \\ can be used within to get better formatting as desired.
% Do not put math or special symbols in the title.
\title{High-Performance Sensorless Control for High-Speed
PMSM with Current Source Inverters}

% author names and affiliations
% use a multiple column layout for up to three different
% affiliations
\author{\IEEEauthorblockN{Nail Tosun\dag, Devinda Molligoda, Xu Deng, and~Barrie C. Mecrow }

\IEEEauthorblockA{Electrical Power Group,\\
Newcastle University\\
Newcastle upon Tyne, NE1 7UR United Kingdom\\
\dag e-mail: n.tosun2@newcastle.ac.uk}
\thanks{This work was supported by the Engineering and Physical
Sciences Research Council (grand number EP/S024069/1).
This work conducted within the Centre for Doctoral Training
in Sustainable Electric Propulsion.}
}

% make the title area
\maketitle

% As a general rule, do not put math, special symbols or citations
% in the abstract
\begin{abstract}
    Current source inverters (CSIs) are emerging as competitive alternatives to voltage source inverters (VSIs) for high-speed motor drives because they inherently suppress harmonic currents, offer output-voltage boosting, and reduce motor-side electromagnetic interference. Sensorless control of CSI-fed permanent magnet synchronous machines (PMSMs) remains less mature than VSI counterparts, primarily because conventional implementations require measurement of all circuit state variables. This paper proposes a weak-resonance-based approximation that reconstructs stator currents from inverter modulation commands and feedforward capacitor-current estimation, thereby reducing the sensing requirement to the dc-link current and two terminal voltages. The proposed observer--PLL structure is analysed for computational complexity and validated through simulations and experiments on a 100~W, 100~krpm GaN-based two-stage CSI prototype (buck + CSI). Quantitative results demonstrate accurate rotor-position tracking and high-bandwidth operation while maintaining the reduced sensor set.
    Proposed method performs well in simulation and experiment, achieving a position estimation error below $5^\circ$ at 100~krpm.
\end{abstract}

% no keywords

% For peer review papers, you can put extra information on the cover
% page as needed:
% \ifCLASSOPTIONpeerreview
% \begin{center} \bfseries EDICS Category: 3-BBND \end{center}
% \fi
%
% For peerreview papers, this IEEEtran command inserts a page break and
% creates the second title. It will be ignored for other modes.
\IEEEpeerreviewmaketitle

\section{Introduction}
Current source inverters (CSIs) have regained attention in motor drives due to the advent of wide-bandgap devices. They offer several advantages over voltage source inverters (VSIs) \cite{10915724}, including inherent short-circuit protection \cite{Jahns2024}, improved dc-link utilisation, improved high-speed performance \cite{8939276, 11193815}, and reduced electromagnetic interference \cite{10476496}. 
However, sensorless control of CSI-fed permanent magnet synchronous machines (PMSMs) remains less mature than its VSI counterpart. Torres et al. proposed a sensorless control method based on high-frequency signal injection \cite{10758742, 9236292, 8921573}; however, this approach requires machine saliency, which is limited in surface-mount PMSMs. 
Prior work \cite{bahramifard2025sensorlessfieldorientedcontrol,9589803,directvoltagemeasurement} has investigated back-EMF-observer-based sensorless control for CSI-fed PMSM drives. However, these approaches typically require measurement of all state variables, including filter-capacitor voltages and stator currents, increasing cost and complexity.

In CSI-fed PMSM drives, output filter capacitor is placed to filter out current harmonics. This capacitor bank forms a parallel $LC$ resonance with the machine inductance. 
There is a potential for this resonance to be excited by switching harmonics or operating-point transients, leading to oscillatory terminal voltages and currents, increased losses, and degraded control performance.
Consequently, conventional approaches often rely on active damping and additional sensing to ensure stability \cite{6241436}. This feature makes CSI-fed drives inherently more complex than VSI-fed PMSM drives.
In order to make CSI-fed drives more competitive, a more efficient sensorless control method is needed.
This paper proposes a high-performance sensorless control method for CSI-fed PMSM drives. 
The proposed method does not require the stator current to be measured, but instead uses the modulation commands and a feedforward capacitor-current estimate.
Elevated switching frequency and a small machine inductance enables this estimation, thus only the dc-link current and two terminal voltages are required. Faster control-loop execution 
can be achieved by this due to the reduced computational burden.

\begin{figure} 
    \centering
    \includegraphics[width=1.0\columnwidth]{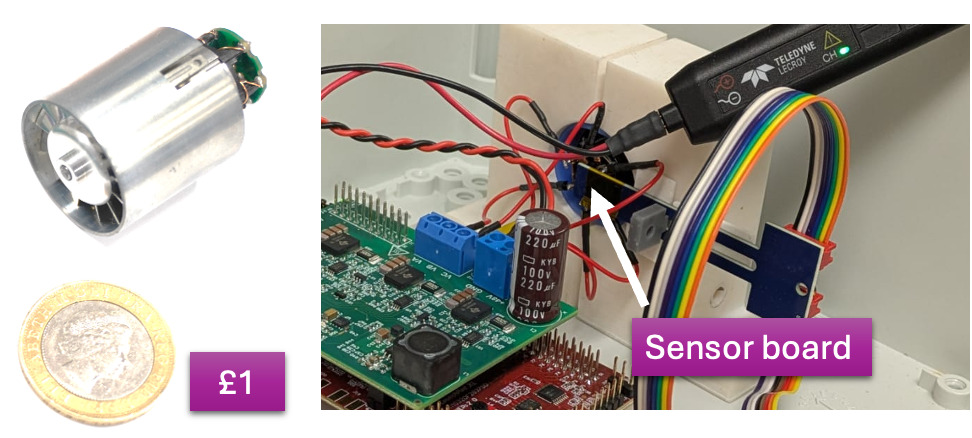}
    \caption{100~W, 100~krpm motor and its relative size compared with a \pounds 1 coin (left). Position sensor attachment (MA600A, Monolithic Power Systems) and its size compared with the machine.}
    \label{fig:introduction}
\end{figure}
A 100~W, 100~krpm PMSM is selected for experimental validation, where the high fundamental frequency demands high-performance sensorless control. The short shaft and limited space make mechanical sensor integration impractical, as shown in Fig.~1. A two-stage GaN-based CSI prototype — comprising a buck converter followed by a CSI power stage — is used. The drive starts via V/f from standstill, with the sensorless algorithm enabled once back-EMF is sufficient for reliable estimation. Simulation and experimental results confirm a position estimation error below $5^\circ$
 at 100~krpm.

\section{Current Source Inverter-Fed PMSM Drives}
CSI-fed PMSM drives are higher-order than VSI drives due to the output filter capacitor. Fig.~\ref{fig:csi} shows the drive topology; the open-loop plant from modulation current to stator current can be modelled as:

\begin{figure} []
    \centering
    \includegraphics[width=1.0\columnwidth]{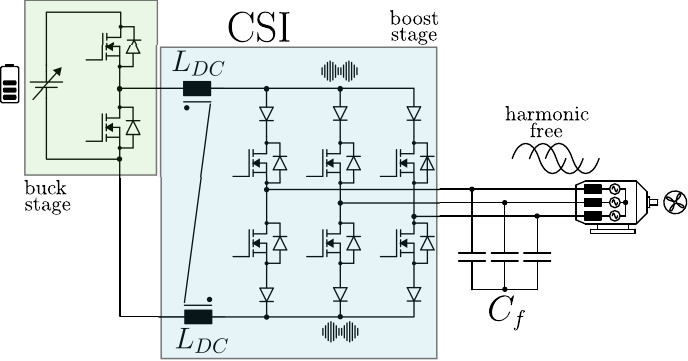}
\caption{Current-source-inverter-fed PMSM drive.}
    \label{fig:csi}
\end{figure}

\begin{equation}
  G(s)=\frac{I_{dq}^{s}(s)}{I_{dq}^{w}(s)}=\frac{1}{s^2L_sC_f+sR_sC_f+1}
\label{eq:csi_transfer_function}
\end{equation}
where $I_{dq}^{s}$ is the stator current, $I_{dq}^{w}$ is the modulation current, $L_s$ and $R_s$ are stator parameters, and $C_f$ is the filter capacitor. The corresponding resonance and damping are:
\begin{equation}
  f_{res} = \frac{1}{2\pi\sqrt{L_s C_f}}
\label{eq:resonant_frequency}
\end{equation}
\begin{equation}
  \zeta = \frac{R_s}{2} \sqrt{\frac{C_f}{L_s}}
\label{eq:damping_ratio}
\end{equation}
High $Q=\frac{1}{2\zeta}$ yields a pronounced resonance, increasing high-frequency losses and often motivating active damping (typically at the cost of extra sensing and inner-loop complexity).

Low-inductance machines reduce $\tau_s=L_s/R_s$, pushing VSI drives toward higher loop bandwidth.
However, smaller $L_s$ lowers $Q$ in CSI drives. With sufficiently large $R_s$ and high switching frequency, the $LC$ resonance becomes weak and less likely to be excited.

\subsection{Conventional CSI PMSM Drive}
\begin{figure}[h!]
    \centering
    \includegraphics[width=1.0\columnwidth]{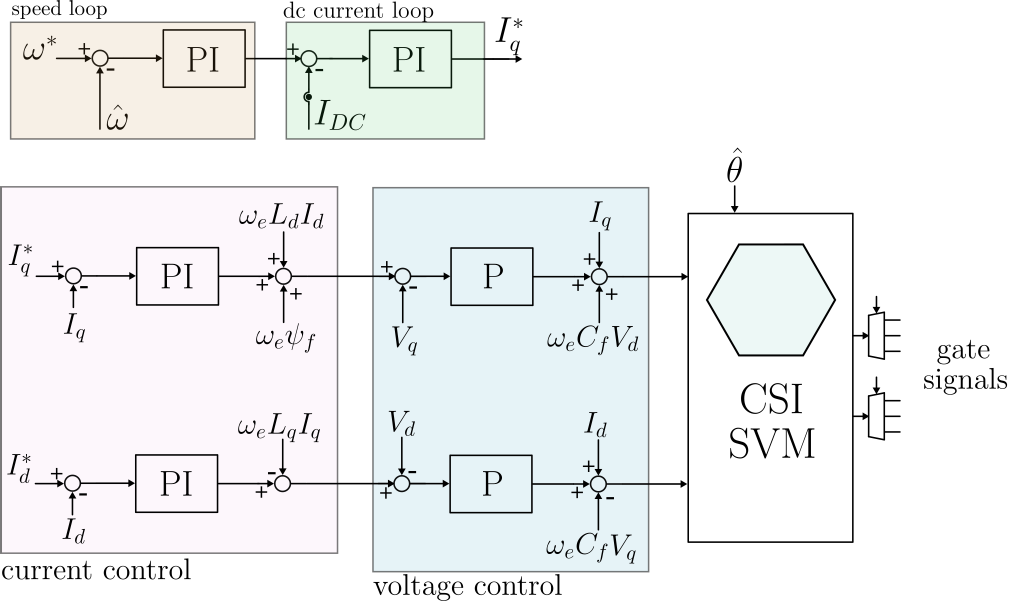}
    \caption{Block diagram of the conventional CSI PMSM drive. }
    \label{fig:conventionalcsi}
\end{figure}

Conventional CSI PMSM drives are commonly implemented as two-stage systems (buck + CSI) because the CSI stage exhibits boost behaviour. The overall control structure is illustrated in Fig.~\ref{fig:conventionalcsi}.
More detail about controller design can be found in \cite{6241436, 10378654,9154581}. CSI PMSM drive control inherently higher order and more complicated than VSI.

\subsection{Weak Resonance Approximation}
If the $L_s$ is sufficiently small and/or $R_s$ is large, the machine impedance can be approximated as resistive, especially
at lower frequencies. In that case, (\ref{eq:csi_transfer_function}) can be approximated as:
\begin{equation}
  G(s) \approx \frac{1}{sR_sC_f + 1}
\label{eq:csi_transfer_function_approx}
\end{equation}

\begin{figure}
    \centering
    \includegraphics[width=1.0\columnwidth]{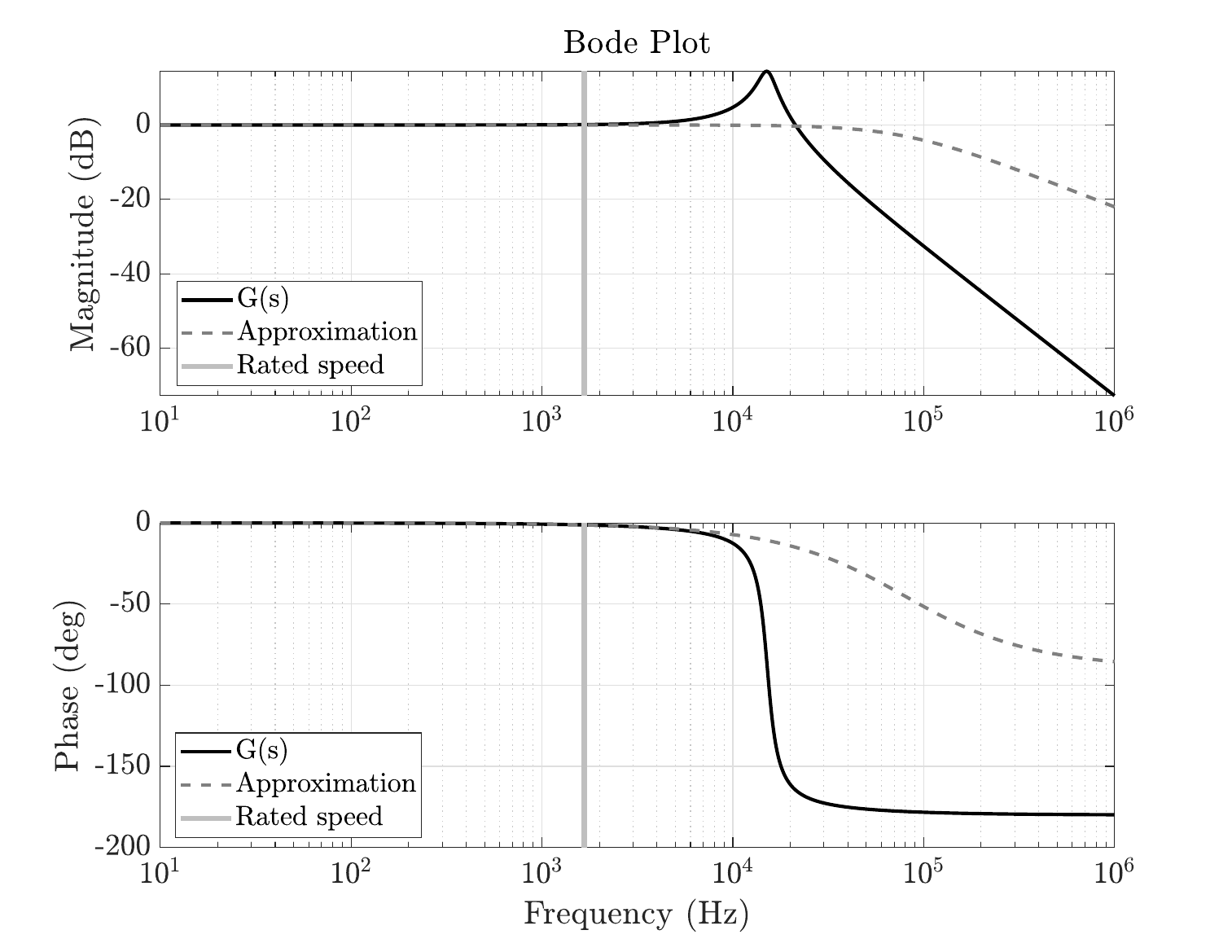}
    \caption{Bode plot of the CSI PMSM drive and its weak resonance approximation.} 
    \label{fig:bodeplot}
\end{figure}

Neglecting the second-order term yields a first-order approximation. The stator current can be adjusted via the modulation current command, while the capacitor current can be compensated using a feedforward term; under this model, output-voltage and dc-link current measurements are sufficient, avoiding direct stator current sensing. Fig.~\ref{fig:bodeplot} compares the full plant and the approximation.

\begin{figure}
    \centering
    \includegraphics[width=1.0\columnwidth]{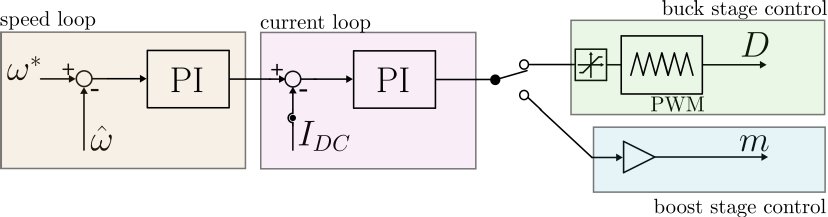}
    \caption{Control loops of the weak resonance CSI PMSM drive.}
    \label{fig:controlloops}  
\end{figure}

The outer-loop structure remains similar to the conventional drive (Fig.~\ref{fig:controlloops}); the dc-link current loop can use duty ratio $D$ and modulation index $m$ depending on operating point. The key input--output relations are:
\begin{equation}
    \frac{V_{ll}}{V_{dc}} = \frac{2}{\sqrt{3}} \frac{D}{m}  
    \label{voltagegain}
  \end{equation}
  
  \begin{equation}
    I_{o} = m I_{dc}
    \label{currentgain}
  \end{equation}

In the $dq$ frame, inverter current is the sum of stator and capacitor currents:

\begin{equation}
  \mathbf{i}^{i}_{dq} = \mathbf{i}_{dq} + \mathbf{i}^{c}_{dq}
\label{eq:csi_current_dq}
\end{equation}

\begin{figure}[h!]
\centering
\includegraphics[width=1.0\columnwidth]{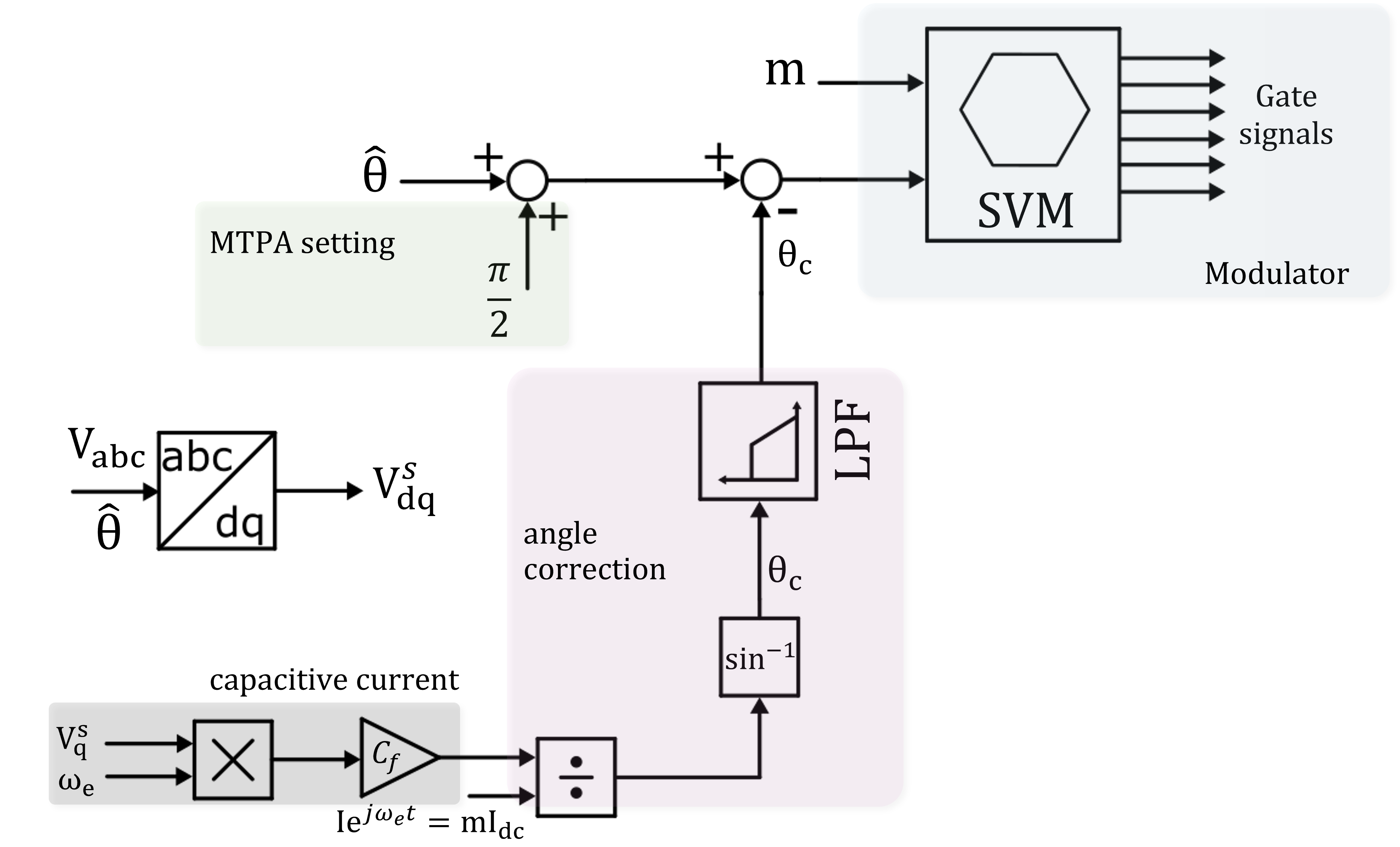}
\caption{Block diagram of the conventional CSI PMSM drive with weak resonance approximation.}
\label{fig:controlweakresonance}
\end{figure}

To improve stator-current tracking, $\mathbf{i}^{c}_{dq}$ can be compensated via a feedforward angle $\theta_c$ (with $V_q$ low-pass filtered). The resulting structure is shown in Fig.~\ref{fig:controlweakresonance}.
\begin{equation}
 \theta_c = \mathrm{sin}^{-1} \left(\frac{\omega_e C_f V_q }{m I_{dc}}\right)
 \label{compensation angle}
\end{equation}

Fig.~\ref{fig:controlweakresonance} shows the current/voltage loop structure under the weak-resonance model; the sensorless estimation is introduced in the next section.
\section{Proposed Sensorless Control Strategy}
The proposed sensorless scheme consists of a back-EMF observer and a phase-locked loop (PLL). For surface-mount PMSMs, back-EMF-based methods are preferred over high-frequency signal injection because saliency is limited.
\begin{table*}
    \centering
    \caption{Comparison of sensorless control methods for CSI/CSC-fed motor drives.}
    \label{tab:csi-sensorless-literature}
    \resizebox{\textwidth}{!}{%
    \begin{tabular}{l l c l l c c}
        \toprule
        \textbf{Reference} & \textbf{Method} & \textbf{Saliency req.} & \textbf{Measurements} & \textbf{Computation} & \textbf{$f_\text{sw}$ / $f_\text{ISR}$} & \textbf{Max.\ speed} \\
        \midrule
        \cite{9589803}
            & PLL
            & No
            & $v_{abc}$, $i_{abc}$, $i_{dc}$
            & PLL
            & --
            & -- \\
        \cite{directvoltagemeasurement}
            & PLL + observer
            & No
            & $v_{abc}$, $i_{abc}$, $i_{dc}$
            & PLL, back-EMF observer
            & --
            & -- \\
        \cite{10758742}
            & HFI (PVE)
            & Yes$^{\dagger}$
            & $v_{C}$, $i_{abc}$, $i_{dc}$
            & BPF, signal demodulation
            & --
            & -- \\
        \cite{high-frequency-injection-CSI-IPM}
            & HFI (pulsating)
            & Yes
            & $v_{C}$, $i_{abc}$, $i_{dc}$
            & BPF, MS-SVM, demodulation
            & 540--720~Hz
            & Zero/low speed \\
        \cite{observer-based-sensorless-control-CSI-PMSM}
            & Back-EMF observer
            & No
            & $v_{C}$, $i_{abc}$, $i_{dc}$
            & Observer, I-f startup
            & --
            & -- \\
        \cite{observer-based-sensorless-control-CSI-PMSM-2}
            & Discrete SMO
            & No
            & $v_{C}$, $i_{abc}$, $i_{dc}$
            & SMO, LPF, DVC
            & 420--540~Hz
            & -- \\
        \cite{observer-based-sensorless-control-CSI-PMSM-3}
            & Exact discrete SMO
            & No
            & $v_{C}$, $i_{abc}$, $i_{dc}$
            & SMO, adaptive filter, DVC
            & 420--540~Hz
            & -- \\
        \midrule
        \textbf{Proposed}
            & \textbf{SMO + PLL}
            & \textbf{No}
            & $\mathbf{v}_{abc}$, $\mathbf{i}_{dc}$
            & \textbf{SMO, PLL}
            & \textbf{200~kHz}
            & \textbf{100~krpm} \\
        \bottomrule
        \multicolumn{7}{l}{\footnotesize $v_{abc}$: motor terminal voltages, $i_{abc}$: motor phase currents, $v_{C}$: filter cap.\ voltages, $i_{dc}$: DC-link current.} \\
        \multicolumn{7}{l}{\footnotesize $^{\dagger}$Works with low-saliency SPM but saliency is still fundamentally required.}
    \end{tabular}}
\end{table*}
Table~\ref{tab:csi-sensorless-literature} compares the proposed method against existing sensorless control approaches for CSI-fed motor drives. The key advantage of the proposed method is the reduced sensor requirement: only the DC-link current and two terminal voltages are needed, eliminating the filter capacitor voltage and stator current sensors required by prior work. This is achieved through the weak-resonance approximation, which allows stator currents to be reconstructed from the modulation commands. Furthermore, the proposed method operates at a switching frequency of 200~kHz enabled by GaN devices, which is significantly higher than prior CSI sensorless implementations, allowing high-speed operation up to 100~krpm.

 \subsection{Back-EMF Observer Design}

The back-EMF in the $\alpha\beta$ frame is:
\begin{equation}
  \mathbf{e_{\alpha\beta}} = 
  \begin{bmatrix}
    e_{\alpha} \\
    e_{\beta}
  \end{bmatrix} = \omega_e \psi_{PM} 
  \begin{bmatrix}
    -\sin(\theta_e) \\
    \cos(\theta_e)
  \end{bmatrix}
\label{eq:backemf}
\end{equation}
where $\omega_e$ is the electrical angular speed, $\psi_{PM}$ is the permanent magnet flux linkage, and $\theta_e$ is the rotor angle.

The stator voltage equation can be written as:
\begin{equation}
  \mathbf{v}_{\alpha\beta} = R_s \mathbf{i}_{\alpha\beta} + L_s \frac{d\mathbf{i}_{\alpha\beta}}{dt} + \mathbf{e}_{\alpha\beta}
\label{eq:stator_voltage_compact}
\end{equation}

From (\ref{eq:stator_voltage_compact}), the back-EMF estimate is:
\begin{equation}
  \hat{\mathbf{e}}_{\alpha\beta} = \mathbf{v}_{\alpha\beta} - R_s \mathbf{i}_{\alpha\beta} - L_s \frac{d\mathbf{i}_{\alpha\beta}}{dt}
\label{eq:backemf_observer}
\end{equation}

(\ref{eq:backemf_observer}) is needed to be solved, both for VSI and CSI drives.  In VSI case, the stator current is measured and terminal voltage is known since it is modulated.
In CSI case inverter current is modulated, thus both terminal voltage and stator currents are needed to measured to solve (\ref{eq:backemf_observer}).
However, the proposed method utilizes the stator current estimation enabled by the weak resonance approximation as it is explained in the previous section.

\subsection{Back-EMF Observer Implementation}
A sliding-mode observer is adopted for robustness \cite{6226863}. The sliding surface is the voltage estimation error:
\begin{equation}
    S(X) = \mbf{v}_{\ab} - \mbf{\hat{v}}_{\ab}
\end{equation}
The Lyapunov candidate function is chosen as $V(X) = \frac{1}{2} S(X)^T S(X)$ which satisfies $V(X)\geq 0$. For stability, $\dot{V}(X)\leq 0$ is required; choosing $K>0$ enforces the sliding condition and thus convergence.
The complete block diagram structure is:
\begin{align}
    \boldsymbol{i}_{\alpha\beta}[k] &= \boldsymbol{i^{i}}_{\alpha\beta}[k] - C \cdot \frac{\boldsymbol{v}_{\alpha\beta}[k] - \boldsymbol{v}_{\alpha\beta}[k-1]}{T_s} \\
    \boldsymbol{\dot{i}}_{\alpha\beta}[k] &= \frac{\boldsymbol{i^{i}}_{\alpha\beta}[k] - \boldsymbol{i^{i}}_{\alpha\beta}[k-1]}{T_s} \nonumber\\
    &\quad - C \cdot \frac{\boldsymbol{v}_{\alpha\beta}[k] - 2\boldsymbol{v}_{\alpha\beta}[k-1] + \boldsymbol{v}_{\alpha\beta}[k-2]}{T_s^2} \\
    \boldsymbol{\hat{v}}_{\alpha\beta}[k] &= R\boldsymbol{i}_{\alpha\beta}[k] + L\boldsymbol{\dot{i}}_{\alpha\beta}[k] + \boldsymbol{\hat{e}}_{\alpha\beta}[k-1] \\
    S[k] &= \boldsymbol{v}_{\alpha\beta}[k] - \boldsymbol{\hat{v}}_{\alpha\beta}[k] \\
    \boldsymbol{\hat{e}}_{\alpha\beta}[k] &= \boldsymbol{\hat{e}}_{\alpha\beta}[k-1] + \frac{K T_s}{L}\text{sgn}(S[k])
\end{align}

where $T_s$ is the sampling time, $[k]$ denotes the current sample, and $[k-1]$, $[k-2]$ denote delayed samples.

\subsection{Phase-Locked Loop Design}
It is well known that observer based position estimation is sensitive to noise and jitter \cite{6698352}. To mitigate this, a PLL is used to track the rotor position. With the small-angle approximation, the phase error can be written as:
\begin{equation}
  \begin{aligned}
  \Delta \theta  &= \hat{e}_{\alpha} \cos(\hat{\theta}_e) - \hat{e}_{\beta} \sin(\hat{\theta}_e) \\
  &= \hat{\omega}_e \psi_{PM}\sin(\theta_e-\hat{\theta}_e)
  \approx \hat{\omega}_e \psi_{PM}(\theta_e-\hat{\theta}_e)
    \end{aligned}
  \label{eq:pll1}
\end{equation}
This linearization enables a standard PI design with open-loop transfer function:
\begin{equation}
  G_{PLL}(s) = \frac{k_p s+ k_i}{s^2}
\label{eq:pll2}
\end{equation}
where $k_p$ and $k_i$ are the proportional and integral gains of the PI controller, respectively. 
The two poles at the origin indicate a type-2 PLL, which tracks a ramp input (position) with zero steady-state error. The PLL also attenuates high-frequency noise and jitter that can arise from direct $\arctan(\cdot)$ angle extraction in the back-EMF observer.

\section{Simulation Results}

\begin{figure*}[t!]
    \centering
\subfigure[Position estimation during start-up.]{
    \includegraphics[width=0.48\textwidth]{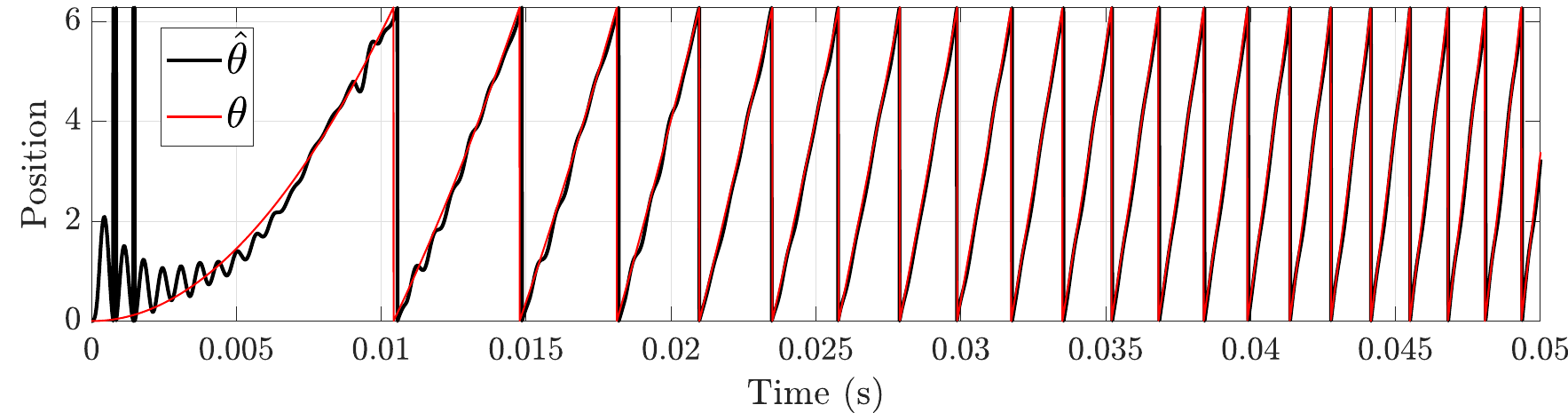}
    \label{fig:position_estimation}
    }
    \hfill
\subfigure[Position estimation error during start-up.]{
    \includegraphics[width=0.48\textwidth]{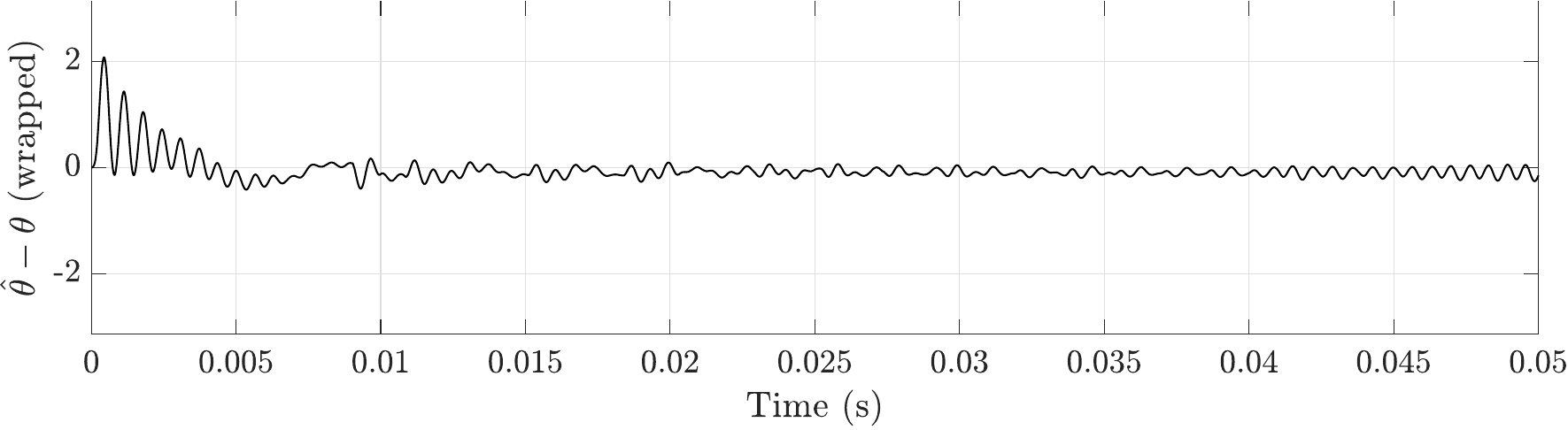}
    \label{fig:position_error}
    }
    \\
\subfigure[Simulation results at 50~krpm.]{
    \includegraphics[width=0.48\textwidth]{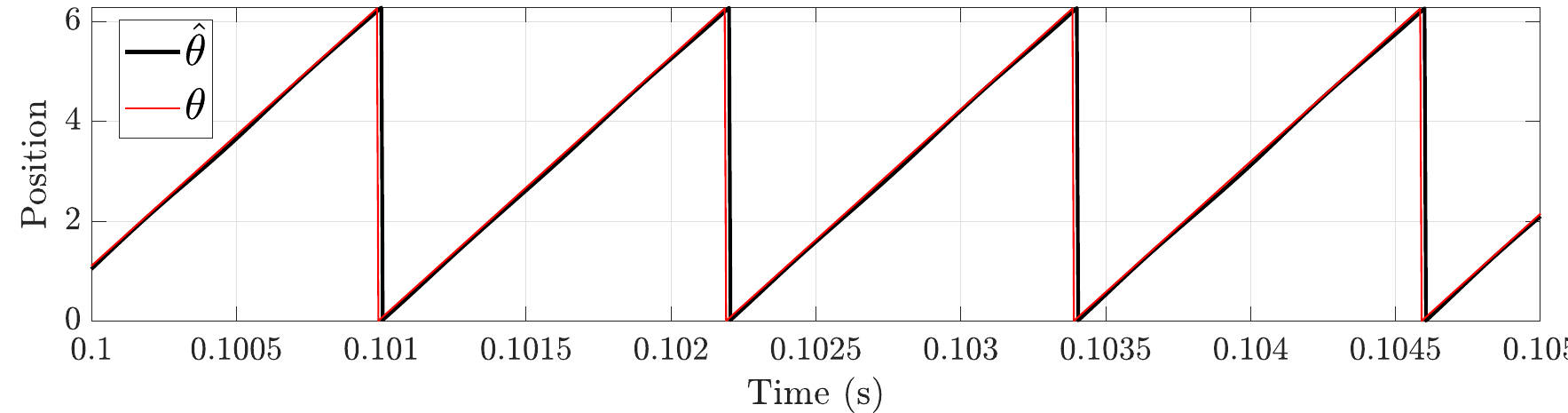}
    \label{fig:sim_50krpm}
    }
    \hfill
\subfigure[Simulation results at 100~krpm.]{
    \includegraphics[width=0.48\textwidth]{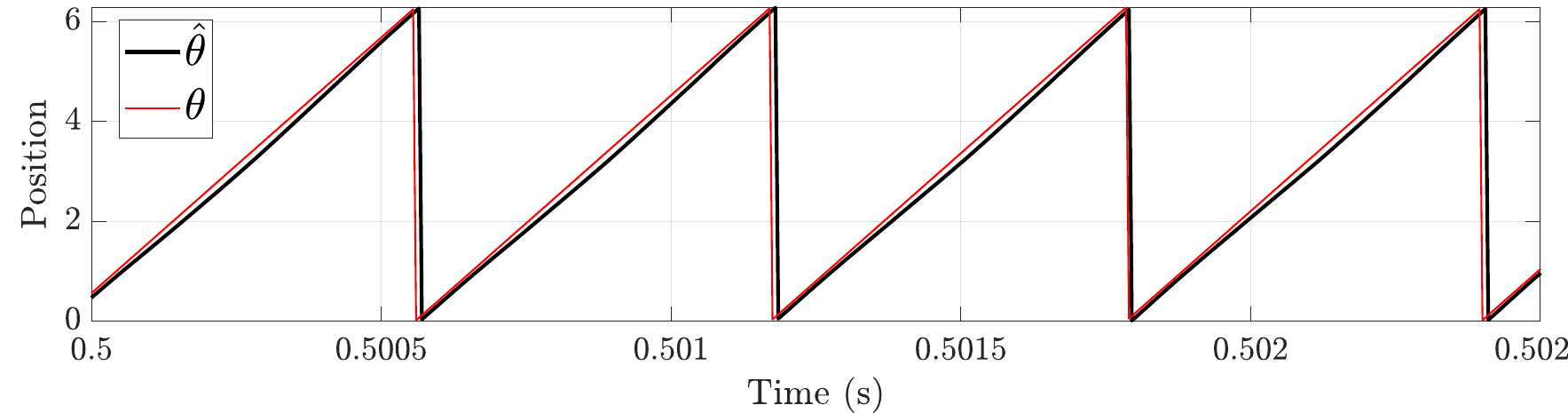}
    \label{fig:sim_100krpm}
    }
    \caption{Simulation results showing the performance of the proposed sensorless control method. (a) Position estimation comparing actual and estimated rotor position during start-up. (b) Position estimation error. (c) Simulation results at 50~krpm showing stator currents and terminal voltages. (d) Simulation results at 100~krpm demonstrating high-speed operation.}
    \label{fig:simulation_results}
\end{figure*}
A PLECS model is developed to validate the proposed sensorless control method using the same machine and system parameters as the experimental setup. This section summarises the parameters and then presents the results.
\subsection{Machine}

A 100~W, 100~krpm high-speed PMSM is selected due to its high fundamental frequency and the resulting need for high-performance sensorless control. 
The machine has very low inductance ($<10\%$ p.u. at rated speed). Fig.~\ref{fig:introduction} shows the prototype and Table~\ref{tab:pmsm_params} lists the parameters.

\begin{table}[!h]
    \caption{PMSM parameters.}
    \centering
    \small
    \renewcommand{\arraystretch}{1.1}
    
    \begin{tabular}{>{\centering\arraybackslash}p{1.3cm} >{\centering\arraybackslash}p{1.6cm} >{\centering\arraybackslash}p{1.7cm} >{\centering\arraybackslash}p{2.5cm}}
    \toprule
    \textbf{Symbol} & \textbf{Value} & \textbf{Unit} & \textbf{Definition} \\
    
    $L_s$ & 23 & $\mu$H & Stator inductance \\
    $R_s$ & 0.40 & $\Omega$ & Stator resistance \\
    $\psi_{PM}$ & 1.1$\times$10$^{-3}$ & Wb & PM flux linkage \\
    $\sigma$ & 1 & -- & $L_q/L_d$ ratio \\
    $K_{load}$ & 8.7$\times$10$^{-11}$ & N$\cdot$m$\cdot$s$^2$/rad$^2$ & Load torque constant \\
    $J$ & 3.7$\times$10$^{-8}$ & kg$\cdot$m$^2$ & Rotor inertia \\
    $B_m$ & 6.4$\times$10$^{-8}$ & N$\cdot$m$\cdot$s/rad & Viscous friction \\
    \bottomrule
    \end{tabular}
    
    \label{tab:pmsm_params}
    \end{table}

\subsection{System Parameters}

System parameters are summarised in Table~\ref{tab:system_params}. The key design variables are the filter capacitor $C_f$ and switching frequency $f_{sw}$; as a guideline, the resonance should be well separated from the switching and fundamental frequencies:

\begin{equation}
  f_{sw} \gg f_{res} \gg f_{fund}.
\label{eq:resonance_condition}
\end{equation}

In motor drives applications, $f_{fund}$ varies with speed (and slot harmonics may appear), so $C_f$ must be selected to avoid exciting the resonance across the operating range. Here, $C_f=5~\mu$F is chosen, yielding a weak resonance and near-sinusoidal terminal voltages (enabling simpler filtering in the back-EMF observer). In per-unit terms, $C_f\approx 0.16$~p.u., keeping operation close to unity power factor.

\begin{equation}
  Z_{base} = \frac{V_{ll}^2}{P_{rated}} = 3.26 \Omega
\end{equation}

\begin{equation}
  C_{base} = \frac{1}{\omega_{rated} Z_{base}} = 29.2 \mu F
\end{equation}

\begin{table}[!h]
\caption{System Parameters.}
\centering
\small
\renewcommand{\arraystretch}{1.1}

\begin{tabular}{>{\centering\arraybackslash}p{1.3cm} >{\centering\arraybackslash}p{1.7cm} >{\centering\arraybackslash}p{1.4cm} >{\centering\arraybackslash}p{2.8cm}}
\toprule
\textbf{Symbol} & \textbf{Value} & \textbf{Unit} & \textbf{Definition} \\
$L_s$ (p.u.) & 0.09 & -- & Stator inductance (p.u.) \\
$R_s$ (p.u.) & 0.17 & -- & Stator resistance (p.u.) \\
$V_{dc}$ & 18 & V & dc-link voltage \\
$C_f$ & 5 & $\mu$F & Filter capacitor \\
$f_{res}$ & 14.8 & kHz & Resonant frequency \\
$f_{sw}$ & 200 & kHz & Switching frequency \\
$f_{fund}$ & 1.66 & kHz & Rated fundamental frequency \\
$Q$ & 0.10 & -- & Quality factor \\
\bottomrule
\end{tabular}

\label{tab:system_params}
\end{table}

\subsection{Simulation Results}
\begin{figure}
    \centering
    \includegraphics[width=1.0\columnwidth]{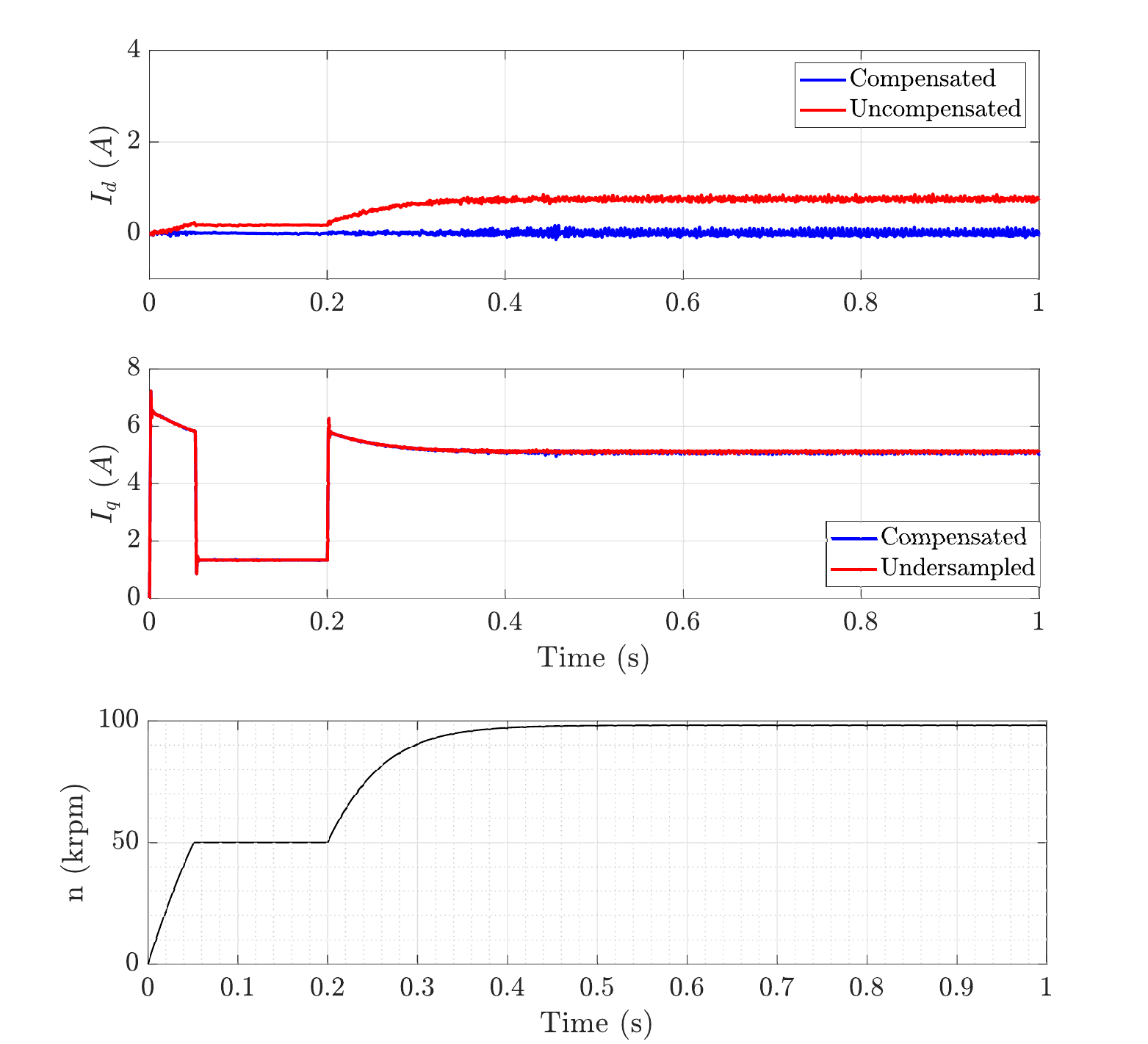}
    \caption{Dynamic response of the proposed sensorless control system. Without compensation, $I_d$ exhibits a steady offset due to capacitor currents.}
    \label{fig:dynamic_response}
\end{figure}

Simulation results using the parameters in Tables~\ref{tab:pmsm_params} and~\ref{tab:system_params} are summarised in Fig.~\ref{fig:simulation_results}. The proposed method tracks rotor position accurately after start-up, with brief transients due to PLL dynamics and low back-EMF. Fig.~\ref{fig:sim_50krpm} and Fig.~\ref{fig:sim_100krpm} demonstrate stable operation at 50~krpm and 100~krpm. Fig.~\ref{fig:dynamic_response} highlights the capacitor-current compensation: without compensation, $I_d$ exhibits a steady offset, whereas compensation drives $I_d \rightarrow 0$.

\section{Experimental Results}
This section presents experimental results for the proposed sensorless control method. The experimental setup is described first, followed by high-speed operating results.

\subsection{Experimental Setup}
\begin{figure}
    \centering
    \includegraphics[width=1.0\columnwidth]{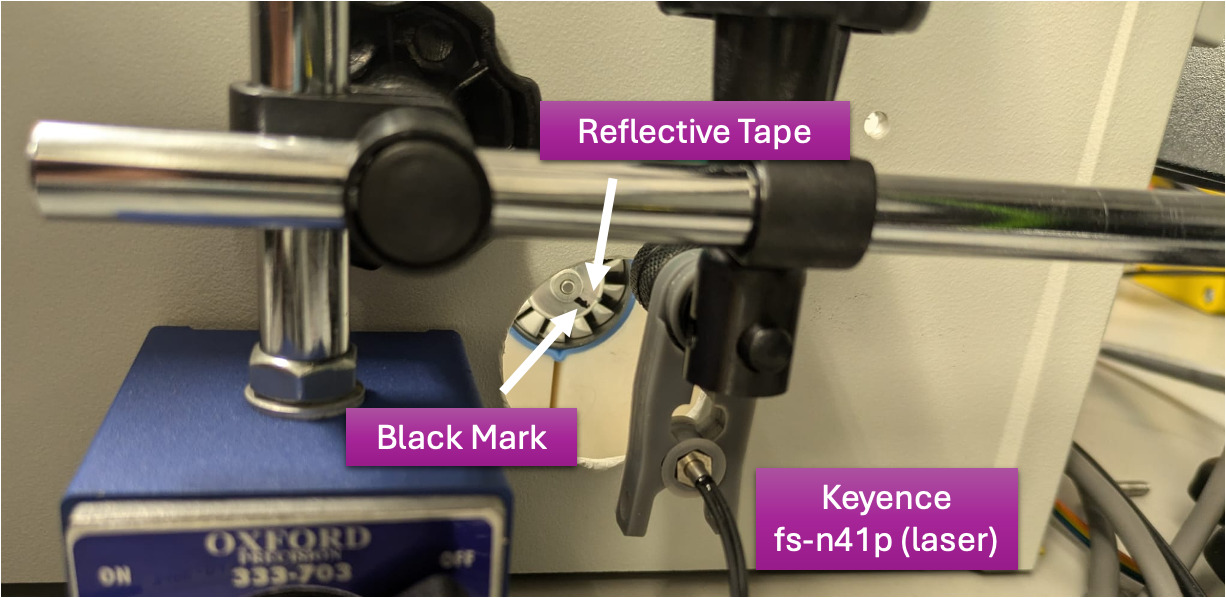}
    \caption{Experimental setup showing the PMSM, laser sensor, reflective tape and black mark.}
    \label{fig:testsetup}
\end{figure}

An accurate position measurement system is required to validate the proposed sensorless control method. However, given the size of the test motor, installing any sensor is challenging. Moreover, conventional position sensors are unreliable at high speeds due to bandwidth limitations.
To overcome these challenges, a laser-based position measurement system is used for the experimental validation. A small reflective tape is attached to the rotor surface, and a laser sensor is used to detect the position. When the laser beam hits the reflective tape, the reflected light is detected by the sensor, and an output pulse is generated. This pulse is used to mark the zero position for reference. 
This is illustrated in Fig.~\ref{fig:testsetup}.

\subsection{High-Speed Operating Results}
\begin{figure*}[ht!]
    \centering
    \includegraphics[width=2.0\columnwidth]{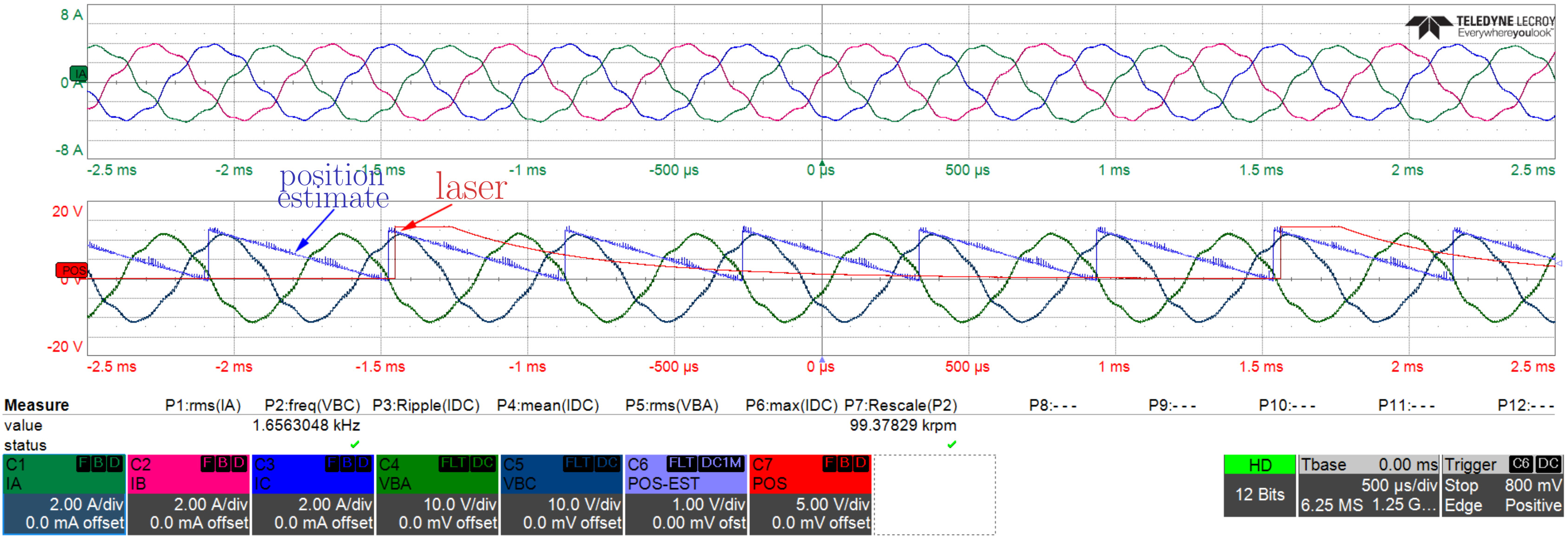}
    \caption{Experimental results demonstrate sensorless control performance at 100~krpm. The upper traces show three-phase currents, and the lower traces show two line-to-line voltages. The blue sawtooth is the estimated position exported via the MCU DAC for visualisation, while the red trace is the laser amplifier output marking the zero-position reference.}
    \label{fig:speed_results}
\end{figure*}

High-speed experimental results are shown in Fig.~\ref{fig:speed_results}. The laser-based reference struggles to detect the zero position during transients due to the laser power amplifier time constant; once the laser output settles, the reference is sufficiently accurate for validation. The proposed method achieves robust high-speed operation, with position estimation error below $5^\circ$ at 100~krpm.

\section{Conclusion}

This paper proposes a high performance sensorless control method for CSI-fed PMSM drives. The proposal uses only dc-link current and two terminal voltage measurements since it eliminates stator current measurements. 
This elimination is done by weak resonance approximation, which is enabled by elevated switching frequency and small machine inductance.
A back-EMF observer and a PLL are used to estimate rotor position and speed. 
The approach is validated in simulation and experiments on a 100~W, 100~krpm prototype, demonstrating high-performance operation for low-inductance machines.

% conference papers do not normally have an appendix

% trigger a \newpage just before the given reference
% number - used to balance the columns on the last page
% adjust value as needed - may need to be readjusted if
% the document is modified later
%\IEEEtriggeratref{8}
% The "triggered" command can be changed if desired:
%\IEEEtriggercmd{\enlargethispage{-5in}}

% references section

% can use a bibliography generated by BibTeX as a .bbl file
% BibTeX documentation can be easily obtained at:
% http://mirror.ctan.org/biblio/bibtex/contrib/doc/
% The IEEEtran BibTeX style support page is at:
% http://www.michaelshell.org/tex/ieeetran/bibtex/
%\bibliographystyle{IEEEtran}
% argument is your BibTeX string definitions and bibliography database(s)
%\bibliography{IEEEabrv,../bib/paper}
%
% <OR> manually copy in the resultant .bbl file
% set second argument of \begin to the number of references
% (used to reserve space for the reference number labels box)

\bibliographystyle{IEEEtran}
%\bibliography{IEEEabrv,Bibliography}
\bibliography{ref}

\begin{IEEEbiographynophoto}{Nail Tosun}
received the B.Sc. and M.Sc. degrees from the Department of Electrical and Electronics Engineering, Middle East Technical University (METU), Ankara, Turkiye, in 2019 and 2021, respectively. He is currently pursuing the Ph.D. degree with Newcastle University, Newcastle upon Tyne, U.K.
He is a member of the Sustainable Electric Propulsion CDT at Newcastle University. His current research interests include electromagnetic FEM analysis, multiphysics models, design optimization of electrical machines, magnetic materials, renewable energy, and power electronics. He has a particular focus on high-frequency power electronics for motor drives, including wide-bandgap semiconductor devices such as GaN-based converters, with emphasis on current source inverter topologies. His work also extends to sensorless motor drive techniques for advanced electric propulsion systems. During his time at METU, he contributed to the electromagnetic design and 3D FEA modelling of high-energy electromagnetic launchers (railguns) under the TUFAN project, and investigated winding type alternation strategies for the refurbishment and efficiency improvement of hydro-generators.
He is the recipient of the WEMDCD Alessandro Costabeber Best Student Paper Award.
 \end{IEEEbiographynophoto}

\begin{IEEEbiographynophoto} {Devinda Molligoda}
received the B.Sc. (Hons) in electronics and telecommunication and the M.Sc. degree in electronics and automation from the University of Moratuwa, Moratuwa, Sri Lanka, in 2009 and 2014 respectively. He is currently working toward the Ph.D. degree in electrical and electronic engineering from Nanyang Technological University (NTU), School of Electrical and Electronic Engineering, Singapore.,From 2009 to 2014, he was an Electronic Design Engineer with Excel Technology Lanka, Colombo, Sri Lanka, before joining the Rolls-Royce@NTU Corporate Lab, Singapore, in 2015, as a Research Engineer. His research interests include three-phase power converters for high-power-density applications, their modulation, control and more-electrical aircraft.
 \end{IEEEbiographynophoto}

\begin{IEEEbiographynophoto}
{Xu Deng}
received the B.Eng. and M.Eng. degrees from Nanjing University of Aeronautics and Astronautics, Nanjing, China, in 2010 and 2013, respectively, and the Ph.D. degree from Newcastle University, Newcastle upon Tyne, U.K, in 2017, all in electrical engineering.,Currently, she is with Newcastle University Academic Track (NUAcT) Fellow in Electrical Power of Electrical Power Research Group, School of Engineering, Newcastle University, Newcastle upon Tyne, U.K. Her research interests include integrated motor drives (IMDs) and advanced control techniques for electric motor drives and high-frequency power electronics.
 \end{IEEEbiographynophoto}

\begin{IEEEbiographynophoto}
{Barrie Mecrow}
received the Ph.D. degree from Newcastle University, Newcastle upon Tyne, U.K., in 1987, with a focus on 3-D eddy current computation applied to turbogenerators.,He commenced his career as a Turbo-Generator Design Engineer with NEI Parsons, Newcastle upon Tyne. In 1987, he became a Lecturer, and in 1998 he became a Professor of electrical power engineering with Newcastle University. He is actively involved with industry in aerospace, automotive, and consumer product sectors, who fund a large range of projects. His research interests include fault tolerant drives, high-performance PM machines, and novel switched reluctance drives.
 \end{IEEEbiographynophoto}
% that's all folks
\end{document}